\documentclass[sigconf]{acmart}

\AtBeginDocument{%
  \providecommand\BibTeX{{%
    \normalfont B\kern-0.5em{\scshape i\kern-0.25em b}\kern-0.8em\TeX}}}

\copyrightyear{2020}
\acmYear{2020}
\setcopyright{acmcopyright}\acmConference[KDD '20]{Proceedings of the 26th ACM SIGKDD Conference on Knowledge Discovery and Data Mining USB Stick}{August 23--27, 2020}{Virtual Event, USA}
\acmBooktitle{Proceedings of the 26th ACM SIGKDD Conference on Knowledge Discovery and Data Mining USB Stick (KDD '20), August 23--27, 2020, Virtual Event, USA}
\acmPrice{15.00}
\acmDOI{10.1145/3394486.3403284}
\acmISBN{978-1-4503-7998-4/20/08}

\usepackage{subcaption} 
\usepackage{pgfplots}
\usepackage{multirow}
\usepackage{amsmath}

\usepackage[ruled,linesnumbered]{algorithm2e}

\def\bW{\textbf{W}}

\def\cG{\mathcal{G}}
\def\cV{\mathcal{V}}
\def\cE{\mathcal{E}}

\begin{document}
\fancyhead{}
\title{M2GRL: A Multi-task Multi-view Graph Representation Learning Framework for Web-scale Recommender Systems}


\author{Menghan Wang}
\authornote{Corresponding author}
\affiliation{%
 \institution{Alibaba Group}}
\email{wangmengh@zju.edu.cn}

\author{Yujie Lin}
\affiliation{%
 \institution{Alibaba Group}}
\email{wanrong.lyj@alibaba-inc.com}
 
\author{Guli Lin}
\affiliation{%
 \institution{Alibaba Group}}
\email{guli.lingl@taobao.com}

\author{Keping Yang}
\affiliation{%
 \institution{Alibaba Group}}
\email{shaoyao@taobao.com}

\author{Xiao-Ming Wu}
\affiliation{%
 \institution{Hong Kong Polytechnic University}}
\email{csxmwu@comp.polyu.edu.hk}

\renewcommand{\shortauthors}{Wang, et al.}
\newcommand{\method}{M2GRL}
\begin{abstract}
Combining graph representation learning with multi-view data (side information) for recommendation is a trend in industry. Most existing methods can be categorized as \emph{multi-view representation fusion}; they first build one graph and then integrate multi-view data into a single compact representation for each node in the graph. However, these methods are raising concerns in both engineering and algorithm aspects: 1) multi-view data are abundant and informative in industry and may exceed the capacity of one single vector, and 2) inductive bias may be introduced as multi-view data are often from different distributions. In this paper, we use a \emph{multi-view representation alignment} approach to address this issue.
Particularly, we propose a multi-task multi-view graph representation learning framework (M2GRL) to learn node representations from multi-view graphs for web-scale recommender systems. \method{} constructs one graph for each single-view data, learns multiple separate representations from multiple graphs, and performs alignment to model cross-view relations. \method{} chooses a multi-task learning paradigm to learn intra-view representations and cross-view relations jointly. Besides, \method{} applies homoscedastic uncertainty to adaptively tune the loss weights of tasks during training.
We deploy \method{} at Taobao and train it on 57 billion examples. According to offline metrics and online A/B tests, \method{} significantly outperforms other state-of-the-art algorithms. Further exploration on diversity recommendation in Taobao shows the effectiveness of utilizing multiple representations produced by \method{}, which we argue is a promising direction for various industrial recommendation tasks of different focus.
 
\end{abstract}




\maketitle

\section{Introduction}

Recently, graph-based recommendation algorithms have significantly improved the prediction performance in academia via learning structural relations from graph data. But in industry there still remains many challenges to build a scalable graph-based recommendation algorithm and beat other industrial algorithms, one of which is how to incorporate graph representation learning with side information (e.g., item's price, user's profile). Side information (or multi-view data~\footnote{We use ``multi-view data'' to denote ``side information'' because we argue that ``multi-view data'' is a more general and reasonable term in the industrial scenario. In this paper, ``multi-view data' also includes the user-item rating data.}) depicts different aspects of items (or users) and plays an important role in industrial recommender systems. In Taobao we have billions of items and each item has hundreds of features, and industrial experiences have shown that the huge volume of multi-view data could alleviate the sparsity problem and improve the recommendation performance \cite{wang2018billion,ying2018graph}.

There are mainly two lines of research that explored how to utilize multi-view data in graph representation learning. One line of research is treating multi-view data (except rating data) as the attributes of items, which are then fed as input of graph-based algorithms. The other line of research is constructing a heterogeneous graph with multi-view data, and then applying graph representation learning techniques (e.g., metapath2vec~\cite
{dong2017metapath2vec}) to learn item embeddings. From the perspective of multi-view learning, these two kinds of research can be categorized into \emph{multi-view representation fusion}. That is, data from multiple views are integrated into a single compact representation. In practice, these methods can effectively address the sparsity problem in recommendations \cite{wang2018billion}.

 

\begin{figure}[t]
\centering 
\includegraphics[width=.5\textwidth]{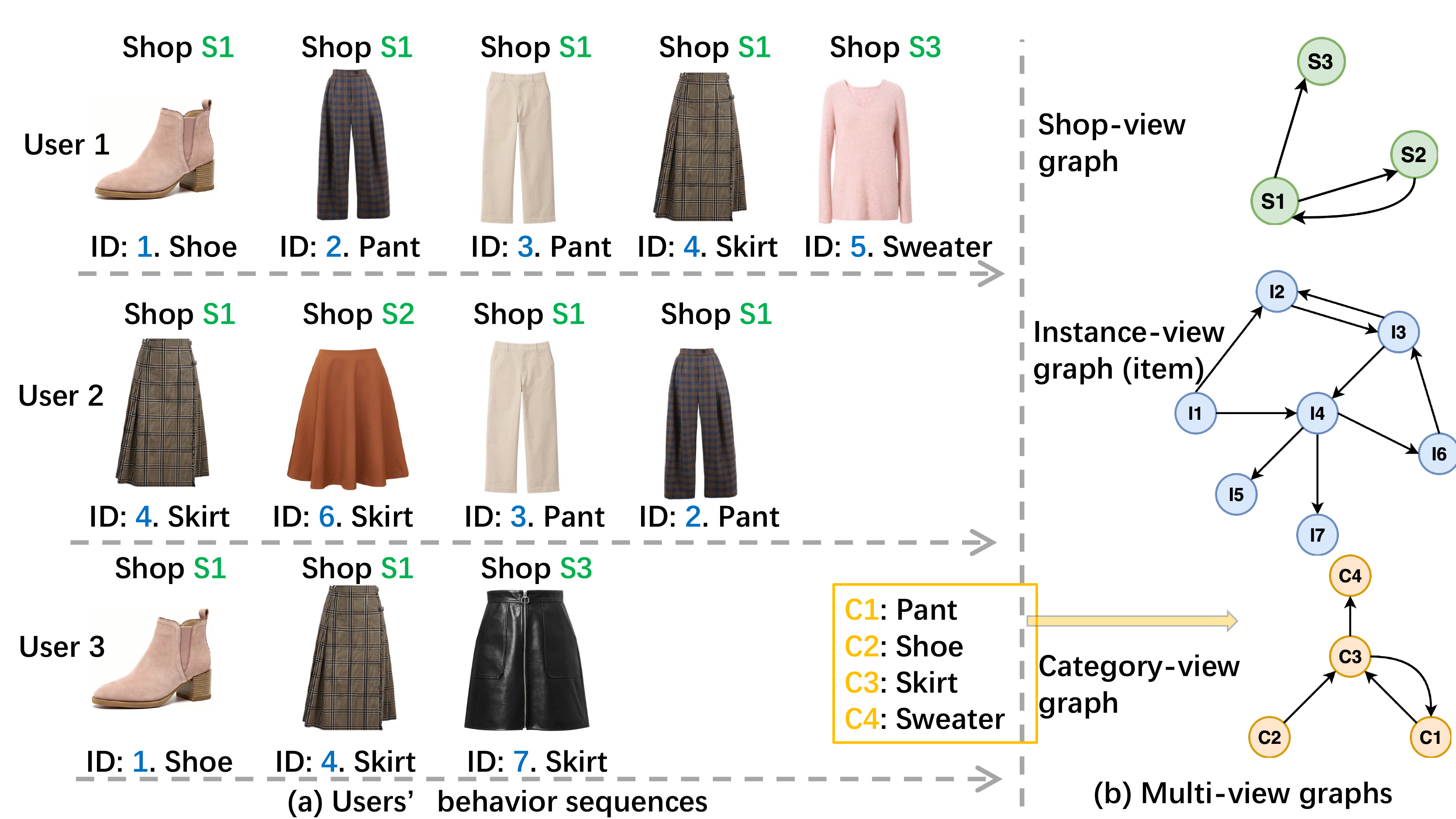}
\caption{An example of users' behavior sequences and the corresponding multi-view graphs. Multi-view graphs constructed from shop, item, and category sequences separately. Different graphs have different structures.}
\label{example}
\end{figure}

However, \emph{multi-view representation fusion} methods raise concerns in both engineering and algorithm aspects when deployed to web-scale recommendation tasks. First, one single vector of representation may lack the capacity to embed multi-view data. In industry there are billions of items and every item may serve multiple interests and needs. But due to engineering concerns, the size of representation vector is often set to $128$ or $256$ even for algorithms based on rating data 
only (single-view), which is already a trade-off between efficiency and accuracy. Fusing multi-view data into one small vector may further sacrifice accuracy. Second, multi-view data may come from different distributions; inductive bias may be introduced if they are not properly handled. For example, as shown in Figure~\ref{example}, we can construct three single-view graphs from users' behavior sequences (self-loops are ignored), i.e., the shop-view graph, the instance-level (item) graph, and the category-level graph. These three graphs have different structures and contain unique information, which are ignored by \emph{multi-view representation fusion} methods. A more proper utilization of multi-view data for graph representation learning is needed in industry.
 
Instead, we argue that a more plausible method is to first learn a separate graph representation for each view of data and then model cross-view relations between different graphs. More concretely, in Figure~\ref{example} we can learn three independent sets of representations for the three graphs respectively, i.e., the green, blue, and yellow graphs. Next, relations of nodes across graphs should be modeled to make the learned representations more reasonable and interpretable. For example, shoe $I1$ (ID: 1) belongs to category $C2$ and is sold in shop $S1$, so the $I1 - C2$ and $I1 - S1$ cross-view relations should influence the final representations of the three nodes.  
The idea of modeling cross-view relations is initially inspired by \emph{multi-view representation alignment}~\cite{li2018survey}, another branch in multi-view learning, which seeks to perform alignment between representations learned from multiple different views. Following this methodology, we can circumvent the two above-mentioned concerns of \emph{multi-view representation fusion} methods: 1) We use multiple representation vectors to represent multi-view data, which can be performed in a distributed manner and hence saves a lot of engineering efforts in configuring the size of a single representation vector as in representation fusion methods. 2) We explicitly preserve local structures of single-view data and model cross-view relations, which is more reasonable and may avoid inductive bias.


Particularly, in this paper, we propose a multi-task multi-view graph representation learning framework (M2GRL) to learn node representations for web-scale recommender systems. \method{} constructs one graph for each single-view data, learns multiple separate representations from multiple graphs, and then performs alignment to model cross-view relations. \method{} comprises two types of tasks: intra-view task and inter-view task. The intra-view task learns the representations of nodes within a single view, while the inter-view task models the cross-view relations between two different graphs. As \method{} tends to have many intra-view and inter-view tasks, we exploit homoscedastic uncertainty~\cite{kendall2017uncertainties} to adaptively tune the loss weights of tasks during training.  The main contributions of this paper can be summarized as follows:
\begin{itemize}
	\item We propose a novel framework \method{} for graph representation learning with multi-view data. To our best knowledge, it is the first work to apply \emph{multi-view representation alignment} in web-scale recommender systems.

\item Our \method{} is scalable, flexible and extensible. It supports unlimited number of views of data, can easily incorporate existing graph representation learning algorithms, and can be distributedly deployed to handle billion-scale data. Besides, the multiple representations learned by \method{} provide item embeddings from different aspects, which can benefit downstream recommendation tasks with different focus. 
	\item Through extensive offline experiments and online A/B tests, we show that \method{} achieves state-of-the-art performance compared to other industrial recommendation algorithms. Further, a use case of diversity recommendation in Taobao is presented to demonstrate the benefits of utilizing the multiple representations learned by \method{}. 
		 
\end{itemize}
The rest of the paper is organized as follows. In Section 2, we introduce related works, including graph representation learning for recommendation and recommendation with multi-view data. Our proposed \method{} and the implementation details are presented in Section 3. We then show offline and online experimental results in Section 4. In Section 5, we describe a use case of diversity recommendation in Taobao with multiple representations produced by \method{}. Finally, we conclude our work in Section 6.

\section{Related Work}
In this section, we discuss the following two lines of research work that are closely related to this paper.

\subsection{Graph Representation Learning for Recommendation}
Graph representation learning aims to learn node or graph embeddings that can capture structural information of graph data. It has become a fast-growing area in the past few years, and many approaches have been proposed. Existing graph representation learning methods could be categorized into three broad categories: 1) Factorization methods (e.g., LINE~\cite{ahmed2013distributed}, NetSMF~\cite{qiu2019netsmf}) aim to approximately factorize the adjacency matrix and preserve structural proximity. 2) Random walk based techniques (e.g., DeepWalk~\cite{perozzi2014deepwalk}, Node2Vec~\cite{grover2016node2vec}, metapath2vec~\cite{dong2017metapath2vec}) use random walks on graphs to obtain node representations; they can be easily deployed in a distributed manner and thus are widely used in industrial applications. 3) Graph convolutional networks (GCNs)~\cite{bruna2013spectral, hamilton2017representation, hamilton2017inductive} perform (approximate) spectral graph convolution to aggregate neighborhood information of each node in a graph. Recent studies have shown superiority of GCNs over factorization methods and random walk based method in small datasets. But GCNs suffer from the efficiency and over-smoothing problems, which prevent their use in industry. 
For recommendation, all the three kinds of methods have been explored to improve the performance of recommender systems. Especially, \citet{wang2018billion} proposed a graph embedding algorithm that first uses random walks to generate samples and then applies word2vec~\cite{mikolov2013efficient} to learn node representations for recommendation at Taobao. \citet{ying2018graph} developed a GCN-based algorithm that combines random walks and graph convolution to generate node embeddings for recommendation at Pinterest.
In this paper, we propose a general framework that allows to incorporate all the three kinds of methods.

\subsection{Recommendation with Multi-view Learning}
In multi-view representation learning, there are two major kinds of methods: 1) multi-view representation fusion, which tries to fuse multi-view data into a single compact representation. 2) multi-view representation alignment, which aims to capture relations among multiple different views through feature alignment. In the literature of recommendation, multi-view data except rating data are collectively called side information, such as temporal information, item's description and users' social network. Currently \emph{multi-view representation fusion} methods are the main stream in recommendation~\cite{zhang2017joint,he2017neural,liang2018variational,wang2018collaborative,zheng2019explore,wang2018modeling}, whereas there also exist some \emph{multi-view representation alignment} methods. For example, \citet{elkahky2015multi} used a deep learning model to map users and items to a latent space where the similarity between users and their preferred items is maximized. \citet{jiang2015deep} introduced a deep cross-modal retrieval method, which considers learning multi-modal embedding from the perspective of optimizing a pairwise ranking problem while enhancing both local alignment and global alignment. However, these works are not designed for graph data, while the focus of this paper is to align node embeddings across different graphs.

\begin{figure*}[th]
\centering 
\includegraphics[height=5cm]{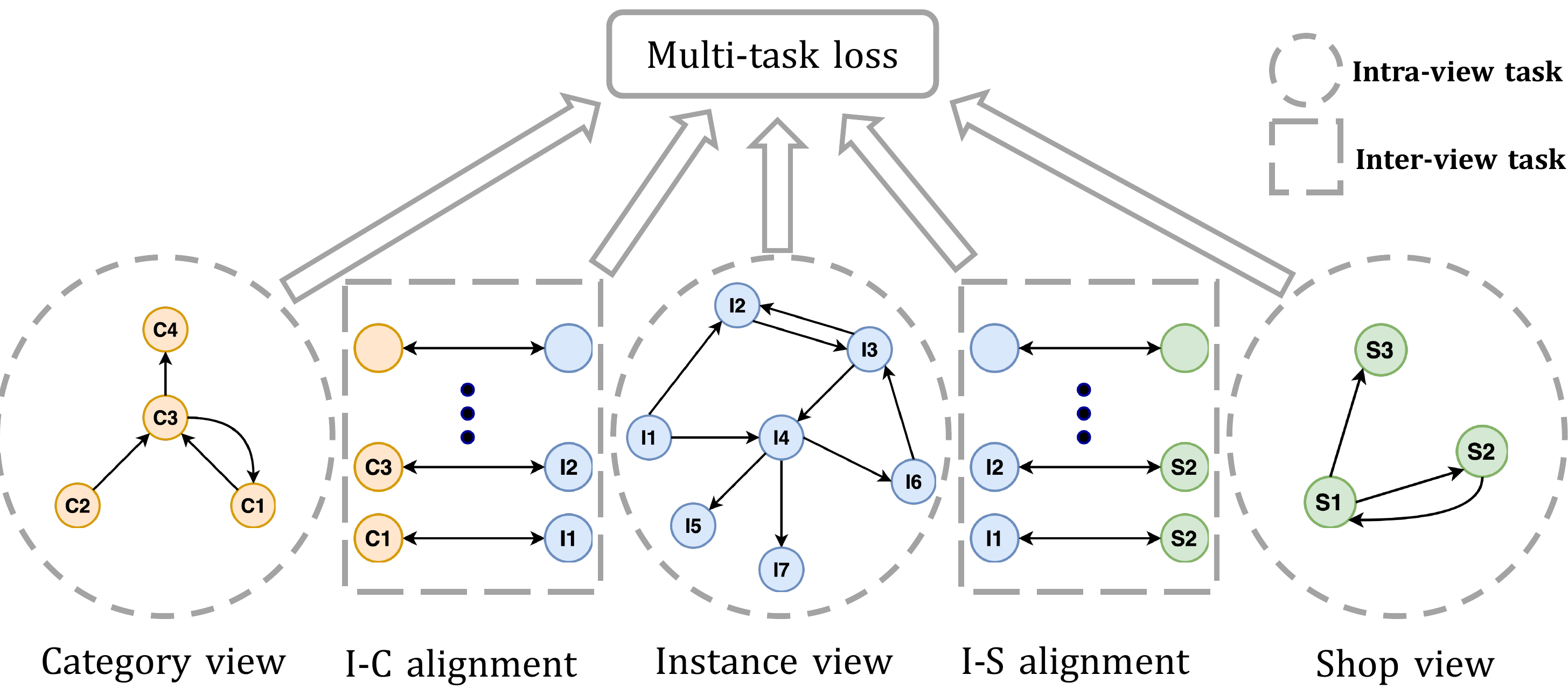}
\caption{Graphical architecture of \method{} with three intra-view tasks and two inter-view tasks.}
\label{mainmodel}
\end{figure*}

\section{\method{} Framework}
In this section, we first introduce the problem setup and how to construct multiple graphs from users' behavior history. Then, we present the overall structure of our \method{} framework and explain its components in detail. 

\subsection{Problem Setup}\label{Sec:prob}
Our task is to learn high-quality representations of items in E-commerce scenarios. These representations can then be used for recommendation by either a nearest-neighbor lookup or various downstream ranking algorithms. Since an item may contain multiple sets of features that reflect different aspects of the item, we can generate multiple representations for each item based on the features. Concretely, we use three-view data in this paper: 1) The instance-view data records the user-item rating data. 2) The category-view data records the category information of items. Category information is a high-level abstraction of items. 3) The shop-view data records the shop information of items. Taobao is a consumer-to-consumer (C2C)  platform with millions of shop selling goods on it. Many shops may sell a same item but with different prices and services. It is worth noting that our framework \method{} can accommodate more views of data. For clarity and simplicity, below we will describe our method with the above three-view data.

\textbf{Graph construction}.
We first construct three graphs including item (instance) graph, category graph, and shop graph for \method{} from users' sequential behaviors. Take the item graph for example, we assume that two items are connected if they occur consecutively in a user's behavior history. The category and shop graphs are constructed similarly. Figure~\ref{example} shows an example of graph construction on three-view data. As we remove consecutive duplicates, the three graphs are of different structures.

\textbf{Node sequence sampling}. A common method to generate node sequences for training samples is random walk and variants. In practice, we generate training samples via extracting sessions from users' behavior history. The main phases are as follows:
\begin{itemize}
	\item \emph{Data Clean}. We noticed that users may click one item and quickly go back to the previous page, which indicates that users are probably not interested in the item. We remove items with duration of stay (after a click) less than two seconds. This operation is also applied in graph construction. 
	\item \emph{Session split and merge}. We extract timestamps from log data that record when a user opens and closes the Taobao App, and use these timestamps to split user behaviors into sessions. Sometimes a session will be hours-long (e.g., running App in the background), so we split a session into two sessions if there is a one-hour idle period. As for session merge, we merge two consecutive sessions if the time span is less than 30 minutes.
\end{itemize}
The category and shop data is bundled with item data, so once an item session is determined we can easily get the corresponding category and shop sessions.

\subsection{Overall Structure}
The model structure of \method{} is illustrated in Figure~\ref{mainmodel}. Particularly, \method{} consists two types of tasks: the intra-view task learning the representations of items within a single-view graph, and the inter-view task modeling the relations of nodes between two different views. We choose five tasks where three tasks are intra-view tasks and the others are inter-view tasks. All the representations generated by the intra-view and inter-view tasks are collectively called multiple representations. 

\subsection{Intra-view Representation Learning}
We treat the intra-view task as a representation learning problem on a homogeneous graph. The intra-view tasks are conditionally independent given the inter-view tasks, so we can apply state-of-the-art graph representation learning methods without much modification. Here we choose the skip-gram model with negative sampling (SGNS), a simple but effective model, to learn node embeddings. SGNS is scalable and can be easily deployed in a distributed manner; many works have shown its effectiveness in extreme large graphs.

We first generate node sequences with the node sequence sampling method described in Section~\ref{Sec:prob}. Then we apply the skip-gram algorithm~\cite{mikolov2013efficient} to learn node embeddings. The training objective of the skip-gram model is to find node representations that are useful for predicting the surrounding nodes in a node sequence. More formally, take the item graph for example, given a sequence of training items $i_1, i_2, i_3, \ldots, i_T$, the objective of the skip-gram model is to maximize the average log probability
\begin{equation}\label{eq:intraloss}
 L_{intra_{i}} = \frac1T\sum_{t=1}^T\sum_{-c\leq j\leq c, j\neq 0} \log p(i_{t+j}|i_t),
\end{equation}
where $c$ is the window size of the context nodes in the sequence. In practice, we find $c=9$ is a good tradeoff between accuracy and training efficiency. The basic skip-gram formulation defines
$p(i_{t+j}|i_t)$ using the softmax function:
\begin{equation}\label{eq:softmax}
p(i_O|i_I) = \frac{\exp\left({v'_{i_O}}^\top v_{i_I}\right)}
{\sum_{w=1}^{W}\exp\left({v'_w}^\top v_{w_I}\right)},
\end{equation}
where $v_w$ and $v'_w$ are the ``Input'' and ``Output'' vector representations
of $w$, and $W$ is the number of items in the item vocabulary.

We use the negative sampling method to approximately maximize the log probability of the softmax function. We first apply negative sampling to generate $k$ negative samples for each positive example, and then Eq.~\eqref{eq:softmax} can be transformed into
 \begin{equation}\label{eq:nce}
   \log \sigma({v'_{i_O}}^\top v_{i_I}) + \sum_{i=1}^k\mathbb E_{i_i\sim
    P_n(i)}\left[\log \sigma(-{v'_{i_i}}^\top v_{i_I})\right],
 \end{equation}
where $P_n(i)$ is the noise distribution for negative sampling, and $\sigma$ is the sigmoid function $\sigma(x) = \frac{1}{1+e^{-x}}$.  
The goal of Eq.~\eqref{eq:nce} is to distinguish the representation of the target item from those of negative samples. The final loss function is Eq.~\eqref{eq:intraloss} with $\log p(i_{t+j}|i_t)$ replaced by Eq.~\eqref{eq:nce}.

\subsection{Inter-view Alignment}
The goal of inter-view tasks is to model the cross-view relations, e.g., \emph{instance - category (I-C)} and \emph{instance - shop (I-S)} relations. If one item $i$ has attribute $x$, we say there is a cross-view relation (\emph{i-x}) between the item and $x$, and vice versa. Instead of directly imposing constraints to two different embedding spaces, we propose an inter-view alignment technique to transform information across views and learn associations of entities in a relational embedding space. Note that one inter-view task is for one type of cross-view relations (e.g., \emph{I-C} relations or \emph{I-S} relations). Specifically, taking the \emph{I-C} alignment task for example, we first map two different embeddings into a relational embedding space via an alignment transformation matrix \bW$_{ic}$. Then, the loss of the inter-view task is formulated as follows:

\begin{equation}
\begin{split}
L_{inter_{i-c}} &=  \sum_{(i,c) \in S(i,c)}  \sigma (W_{ic} \cdot e_{i})^{T} \sigma (W_{ic} \cdot e_{c}) \\
		      &	 - \sum_{(i,c') \notin S(i,c)}  \sigma (W_{ic} \cdot e_{i})^{T} \sigma (W_{ic} \cdot e_{c'}),
\end{split}
\label{eq:intertask}
\end{equation}
where $\sigma$ is the sigmoid activation function, and $c'$ are chosen by negative sampling. The inter-view tasks also generate representations; $\sigma(W_{ic} \cdot e_{i})$ and $ \sigma(W_{ic} \cdot e_{c})$ can be regarded as two representations from different aspects in the relational embedding space.  

\subsection{Learning Task Weights with Homoscedastic\\ Uncertainty }
\method{} is concerned about jointly optimizing multiple related tasks. A na\"ive but popular approach is to define a total loss function that is a linear combination of the loss of each individual task:
 \begin{equation}
\label{eqn:basic_loss}
L_{total}= \sum_i w_i \cdot L_{intra_{i}} + \sum_j w_j \cdot L_{inter_{j}},
\end{equation}
where $\{w_i\}$ and $\{w_j\}$ are hyper-parameters that balance the importance of different losses. However, manually tuning these hyper-parameters is expensive and intractable in web-scale recommendation scenarios. An alternative approach~\cite{xu2018multi} in practice is to optimize each task iteratively, which, however, may be stuck at a local optimum and fail to perform well in some tasks. 

Instead, we exploit the idea of homoscedastic uncertainty~\cite{kendall2017uncertainties} to automatically weigh the loss of each task during model training. Following the work of  \cite{kendall2018multi}, we adjust each task’s relative weight in the total loss function by deriving a multi-task loss function based on maximizing the Gaussian likelihood with task-dependant uncertainty. Note that the intra-view and inter-view tasks are classification tasks.
We rewrite the loss function (Eq.~\eqref{eqn:basic_loss}) as follows:
 \begin{equation}
\label{eqn:new_loss}
\begin{split}
L_{total}(X;\Theta, \Sigma) = \sum_{t}^{T}L_{t}(x_t; \theta_t, \sigma_{t}). 
\end{split}
\end{equation}
where $L_{t}$ is the classification loss function for task $t$, $x_t$ and $\theta_t$ are the data and model parameters for task $t$ (either intra-view task or inter-view task) respectively, and $\sigma_{t}$ is the corresponding task uncertainty.

We represent the likelihood of the model for each task as a scaled version of the model output $f(x_{t})$ with uncertainty $\sigma_t$ squashed by a softmax function:
\begin{equation}
P(C = c|x_t, \theta_t, \sigma_t) = \frac{exp( \frac{1}{\sigma_t^2} f_c(x_t))}{ \sum_{c'=1}^{} exp( \frac{1}{\sigma_t^2} f_{c'}(x_t))},
\end{equation} where $ f_c(x_t)$ is the $c$-th element of the vector $f(x_{t})$.
Using the negative log likelihood, we express the classification loss with uncertainty as follows:
\begin{equation} \label{eg:class_loss}
\begin{split}
L_t(x_t, \theta_t, \sigma_t) & = \sum_{c} - C_c log P(C_c=1| x_t, \theta_t, \sigma_t) \\
& = \sum_{c} - C_c log(exp( \frac{1}{\sigma_t^2} f_c(x_t))) 
							 + log \sum_{c'}  exp( \frac{1}{\sigma_t^2} f_{c'}(x_t)).
\end{split}
\end{equation}
Applying the same assumption in \cite{kendall2018multi}:
\begin{equation}
	 \frac{1}{\sigma^2} \sum_{c'}^{} exp( \frac{1}{\sigma^2} f_{c'}(x_t)) 
	 \approx 
	 (\sum_{c'}^{} exp(f_{c'}(x_t)))^ \frac{1}{\sigma^2},
\end{equation}
Eq.~(\ref{eg:class_loss}) can be simplied as:
\begin{equation}\label{eq:loss_approx}
\begin{split}
L_t(x_t, \theta_t, \sigma_t) \approx \frac{1}{\sigma_t^2} \sum_{c} - C_c log P(C_c=1| x_t, \theta_t) + log(\sigma_t^2).
\end{split}
\end{equation}
We use Eq.~\eqref{eq:loss_approx} to approximate the loss of each task $t$ in Eq.~\eqref{eqn:new_loss} and obtain the final multi-task loss. $\sigma_t$ can be interpreted as the relative weight of the loss of task $t$. When $\sigma_t$ increases, the corresponding weight decreases. Additionally, $log(\sigma_t^2)$ serves as a regularizer to avoid overfitting. For numerical stability, we trained the network to predict $log(\sigma_t^2)$ instead of $\sigma_t^2$. All network parameters and the uncertainty task weights are optimized with stochastic gradient descent (SGD).

\begin{table*}[t]
	\centering
	\caption{Datasets Statistics.}
	\label{tb:statistics}
	\begin{tabular}{c|ccc|ccccc}
		\toprule
		Dataset      & \#Item & \#Category  & \#Shop  & \#Item-edges    & \#Cate-edges & \#Shop-edges      & \#Item-Cate links  & \# Item-Shop links     \\ \midrule
		
		Taobao       & $1.0 \times 10^7$ & $1.2 \times 10^{4}$  & $1.3 \times 10^6$   &  $1.8 \times 10^{10}$  &  $3.5 \times 10^8$   &  $4.4 \times 10^{8}$  & $1.5 \times 10^8$   &  $1.4\times 10^8$    \\ \midrule
		Movielens  & 27,278  & 1,128  & -   & 10,463,449  & 1,712,087  & -  & 9,110,188      &  -\\   \bottomrule
	\end{tabular}
\end{table*}

  \begin{table*}[thb]
        \caption{Comparisons of different models on offline datasets.}
        \label{tab:offexp}
        \begin{tabular}{l|c|c|c|c|c|c|c|c}
            \hline
            \multirow{2}{*}{Models} &  \multicolumn{4}{c|}{Taobao}              &                \multicolumn{4}{c}{Movielens}          \\
            \cline{2-9}
                           & HitRate@50 & Recall@50 & Precision@50 & F1@50 & HitRate@50 & Recall@50 & Precision@50 & F1@50 \\ 
            \hline
             LINE           & 9.55\% & 1.96\% & 0.22\% & 0.40\%                 & 24.89\% & 11.08\% & 0.62\% & 1.17\% \\
           DeepWalk           & 31.66\% & 12.11\% & 1.13\% & 2.06\%                 & 34.72\% & 17.43\% & 0.97\% & 1.83\% \\
           Node2Vec          & 31.79\% & 12.20\% & 1.14\% & 2.08\%                & 36.43\% & 18.14\% & 1.01\% & 1.91\% \\
           EGES$_{asy}$          & 32.42\% & 12.45\% &1.17\% & 2.13\%                 & 42.46\% & 22.01\% & 1.23\%  & 2.33\% \\
           GraphSage           & 7.16\% & 1.98\% & 0.19\% & 0.35\%                 & 2.23\% & 0.76\% & 0.05\% & 0.09\% \\
           GRU4Rec     & 27.32\% & 11.45\% & 1.08\% & 1.97\%                 & 30.81\% & 13.39\% & 0.76\% & 1.44\% \\
           YouTube-DNN      & 29.07\% & 11.97\% & 1.11\% & 2.03\%             & 38.27\% & 18.73\% & 1.06\% & 2.01\% \\
            \hline
            M2GRL          & \textbf{33.24\%} & \textbf{12.65\%} & \textbf{1.19\%} & \textbf{2.17\%}                   & \textbf{42.62\%} & \textbf{22.16\%} & \textbf{1.24\%} & \textbf{2.35\%} \\ 
            \hline
        \end{tabular}
    \end{table*}

\section{System Deployment}  
\begin{figure}[t]
\centering 
\includegraphics[height=5.5cm]{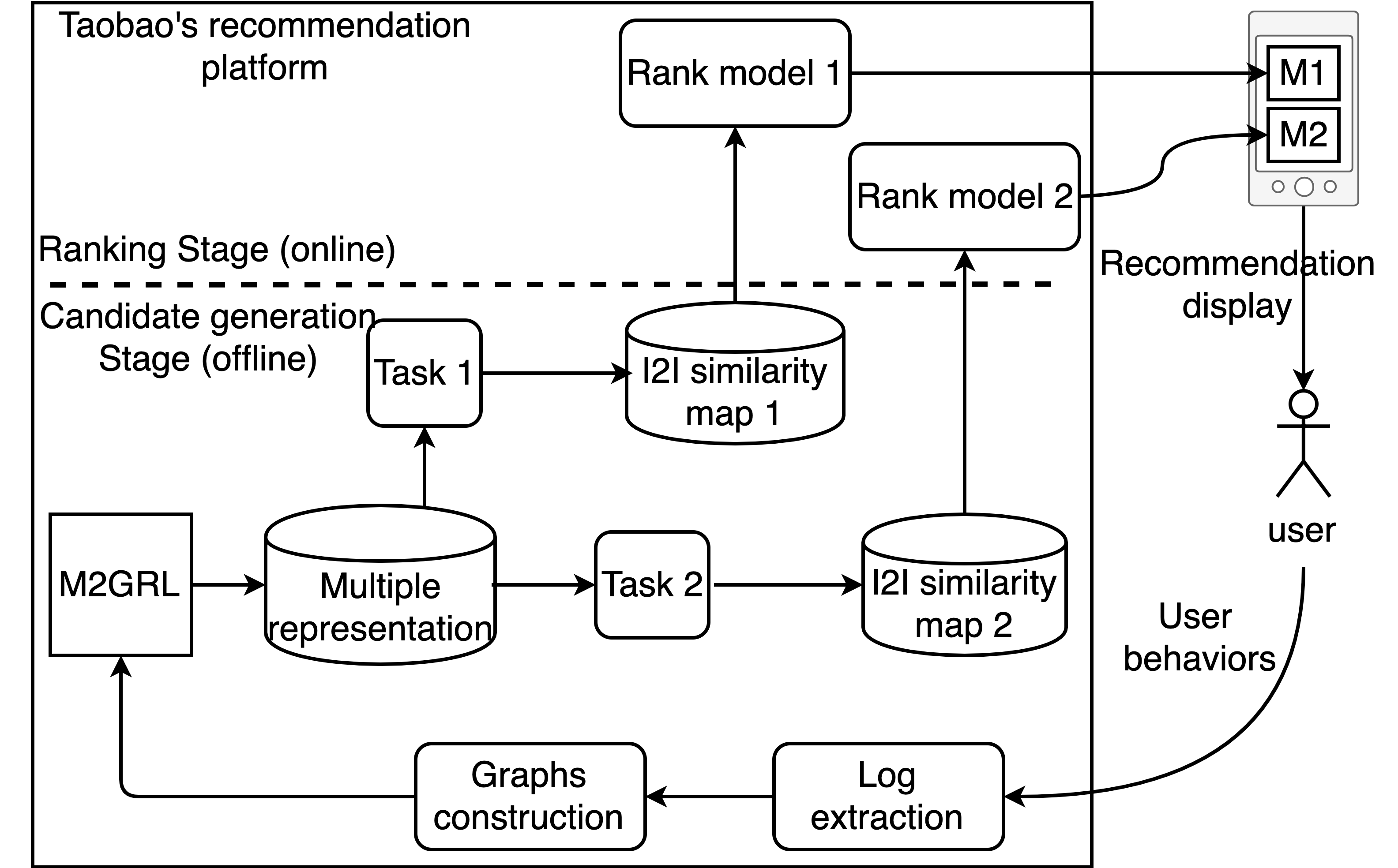}
\caption{Deployment of M2GRL in Taobao's recommendation platform. M1 and M2 denote two modules.}
\label{fig:system_architecture}
\end{figure}

In this section, we introduce the deployment of M2GRL in Taobao's recommendation platform as shown in Figure~\ref{fig:system_architecture}. In the candidate generation stage, the system constructs multiple single-view graphs after extracting multi-view data from raw log data. Then, the system runs M2GRL in a distributed manner to produce multiple representations. In Taobao, we have many recommendation modules (downstream tasks), and we use different strategies (e.g., task 1 and task 2 in Figure~\ref{fig:system_architecture}) to generate \emph{I2I (item-to-item) similarity maps} based on the goal of each module. Given an item as a trigger, items with high similarity scores in the map are chosen as candidates. Note that the size of candidates is much smaller (usually thousands) compared with the size of the corpus (hundreds of millions). In the ranking stage, rank models request \emph{I2I similarity maps} for candidates and use online models to score candidates. Items with top scores are finally recommended and displayed to users.

Training one version of \method{} takes about a dozen hours, so the model is deployed offline in a daily-run mode. We currently support two downstream tasks. One is general recommendation that generates the similarity map using inner product search on the instance-view representations of \method{}. The other is diversity recommendation that first uses the multi-view metric model introduced in Section~\ref{downstream_task} to produce new representations, and then generates the similarity map via inner product search. More downstream tasks with different focus are under development.

\section{Experiments}      
 In this section, we conduct comprehensive experiments to verify the effectiveness of our proposed \method{} framework.

\subsection{Offline Evaluation}
\subsubsection{Datasets and Experimental Settings}
We choose two datasets for evaluating recommendation performance.
One is Taobao, which is extracted from Mobile Taobao App and contains historical behaviors of users on top ten million popular items of Taobao in 3 days (Sep. 02. 2019 to Sep. 05. 2019). The user behaviors on Sep. 06. 2019 is used as test data. The other dataset is Movielens $20$M \footnote{https://grouplens.org/datasets/movielens/}, one of the most widely-used public datasets for recommendation.  
For Taobao, we construct three single-view graphs (i.e., shop, instance, category) according to Section \ref{Sec:prob}. For Movielens, we treat the tag of a movie as its category information and construct single-view graphs similarly. Note that Movielens does not have shop information. So we only construct two graphs for Movielens in \method{}.
The statistics of the two datasets are summarized in Table~\ref{tb:statistics}.

\subsubsection{Baselines}
We evaluate the performance of \method{} against the following state-of-the-art graph-based methods and deep-learning-based methods for learning item embeddings.
\begin{itemize}
   \item\textbf{Deepwalk} \cite{perozzi2014deepwalk}. This approach learns low-dimensional feature
representations for graph nodes by simulating uniform random walks. 
   \item\textbf{LINE} \cite{tang2015line}.  This method learns graph embedding via a customized loss function to preserve the first-order or second-order proximities in a graph separately. 
   \item\textbf{Node2Vec} \cite{grover2016node2vec}. This algorithm treats network embedding learning as a search-based optimization problem. It uses a biased random walk that is a tradeoff between DFS and BFS.
  \item\textbf{EGES$_{asy}$} \cite{wang2018billion}.
  This method incorporates side information into graph embedding via a weighted average layer to aggregate the embeddings of side information. It is a typical \emph{multi-view representation fusion} method. 
  \item\textbf{GraphSage} \cite{hamilton2017inductive}.
  This method learns node embeddings via aggregating information recursively from neighborhood nodes in a graph. It is a variant of graph convolutional networks. 
  \item\textbf{GRU4Rec} \cite{hidasi2015session}.
  A GRU-based RNN model for session-based recommendations, which outperforms traditional methods significantly.
  \item\textbf{YouTube-DNN} \cite{covington2016deep}.
  A recommendation approach proposed by YouTube based on deep neural networks, which is popularly used for industrial recommendation systems. 
  
  \end{itemize}
Hyperparameter tuning is conducted by grid search, and each method is tested with the best hyperparameters for a fair comparison. For \method{}, we generate recommendations via inner product search on the instance-view (item) representations.

\subsubsection{Offline metrics}
To evaluate the offline performance of different methods, we use \textbf{Precision@K}, \textbf{Recall@K} , \textbf{HitRate@K} and \textbf{F1@K} metrics (the last two metrics are defined in the Appendix).

Define the recalled set of items for a user $u$ as $P_u$ ($|P_u | = K$) and the user's ground-truth set as $G_u$. Precision@K reflects how many interested items for a user in the candidates. It is calculated as
    \begin{displaymath}
    \mbox{Precision@K($u$)} = \frac{|P_u \cap G_u|}{K}.
    \end{displaymath}
    
    Recall@K represents the coverage in the user's ground-truth set. It is  calculated as
    \begin{displaymath}
    \mbox{Recall@K($u$)} = \frac{|P_u \cap G_u|}{|G_u|}.
    \end{displaymath}
   

\subsubsection{Offline Experimental Results}
We show the experimental results in Table \ref{tab:offexp}, from which we can find that \method{} outperforms other baselines consistently on both datasets. This demonstrates the effectiveness of our proposed method. Further, we can get the following findings. 1) EGES$_{asy}$ is the best baseline. It beats Node2Vec, DeepWalk and LINE that follow the same line of algorithms, mainly due to the utilization of side information. Side information can alleviate the sparsity problem and thus improve the recommendation performance. However, in EGES$_{asy}$, embeddings of side information are directly averaged with item embedding, ignoring the heterogeneities of items and side information. Our method \method{} takes a \emph{multi-view representation alignment} method to address this issue and outperforms EGES$_{asy}$.
2) Node2Vec and DeepWalk outperform LINE because LINE only preserves information of a two-hop neighborhood, while Node2vec and DeepWalk can learn information from nodes that are a variable number of hops away. Moreover, Node2vec uses weighted random walks and is superior to DeepWalk on both the Movielens and Taobao datasets.
3) We notice that GraphSage performs very poorly. One main reason is that the item graph is built on users' sequential behaviors and it is dense and noisy, which may undermine the effectiveness of GraphSage because the training phase of GraphSage aims to learn aggregators rather than node representations. In addition, the shallow structure of GraphSage (two or three layers) also limits the performance of the produced representations. 
4) GRU4Rec and YouTube-DNN are sequential recommender systems, and they are empirically inferior to random-walk-based methods (e.g, DeepWalk and \method{}). Meanwhile, deploying sequential recommendation algorithms to online environment needs extra engineering efforts and computational overheads: the system has to call users' sequential behavior to predict and rank items for every recommendation request. In contrast, \method{} uses a lookup table.
By the above analysis, it can been seen that \method{} is scalable and competitive.

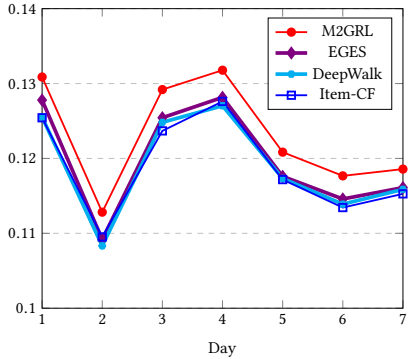
\begin{figure}
\begin{tikzpicture}[scale=0.7]
\begin{axis}[
    xlabel={Day},
    xmin=1, xmax=7,
    ymin=0.1, ymax=0.14,
    xtick={1,2,3,4,5,6,7},
    ytick={0.10,0.11,0.12,0.13,0.14,0.15 },
    legend pos=north east,
    ymajorgrids=true,
    grid style=dashed,
]
 \addplot[
    color=red,
    mark=*,
    line width=1pt,
    ]
    coordinates {
    (1,0.130871429)(2,0.112823351)(3,0.129192881)(4,0.131780325)(5,0.120837163)(6,0.117672143)(7,0.118576179)
    };
    \addlegendentry{M2GRL}
    
   \addplot[
    color=violet,
    mark=diamond,
    line width=2pt,
    ]
    coordinates {
    (1,0.12779027)(2,0.109192631)(3,0.125436271)(4,0.128154746)(5,0.117599)(6,0.114585977)(7,0.116099297)
    };
   \addlegendentry{EGES}
    
\addplot[
    color=cyan,
    mark=asterisk,
    line width=2pt,
    ]
    coordinates {
    (1,0.125523324)(2,0.108330539)(3,0.124821684)(4,0.127030061)(5,0.117324716)(6,0.11389683)(7,0.115867363)
    };
   \addlegendentry{DeepWalk}

\addplot[
    color=blue,
    mark=square,
     line width=1pt,
    ]
    coordinates {
  (1,0.125430192)(2,0.109516976)(3,0.123678807)(4,0.127610752)(5,0.117173062)(6,0.113438693)(7,0.115268844)
    };
     \addlegendentry{Item-CF}
\end{axis}
\end{tikzpicture}
 \caption{ Online CTRs of different methods in 7 days in November 2019}
\label{exp:online}
\end{figure}

\subsection{Online A/B Test}
  
We report the online A/B test experiments of \method{} and other indusitrial algorithms (i.e., EGES, DeepWalk and Item-CF) deployed at Taobao. We choose $150$ million top popular items and use $57$ billion samples for training. \method{} is implemented on PAI-tensorflow (a machine learning platform in Alibaba), and is trained in a distributed manner with $25$ server nodes and $100$ worker nodes. The total amount of memory used in training is 2000GB. The chosen metric is Click-Through-Rate (CTR) on the homepage of Mobile Taobao App. We implement the above four methods and then generate a number of similar items for each item as recommendation candidates. The final recommendation results on the homepage of Taobao is generated by the ranking engine, which is built on a deep neural network model. We use the same method to rank the candidate items in the experiment. Therefore, the recommendation performance, i.e., CTR, can represent the effectiveness of different methods in the matching stage. We deploy the four methods in an A/B test framework and the results of seven days in November 2019 are shown in Figure \ref{exp:online}. Note that the item-based CF method is an enhanced version of the classical Item-CF model; it has been deployed in Taobao for several years and is still running in some tasks. It computes similarities between items according to item co-occurrences and user voting weights. 

From Figure \ref{exp:online}, we can see that M2GRL outperforms EGES, DeepWalk, and Item-CF consistently in terms of CTR, which demonstrates the effectiveness of utilization of muti-view graph embedding. Compared with the best baseline EGES, \method{} achieves a $5.76\%$ relative improvement on average, which is a significant boost in recommendation performance considering that there are hundreds of millions of active users in Taobao everyday. Further, EGES outperforms DeepWalk consistently, which demonstrates the usefulness of side information. The results are similar to what we observe in the offline experiments.

\begin{figure} 
     \centering
     \begin{subfigure}[b]{0.23 \textwidth}
         \centering
         \includegraphics[width=\textwidth]{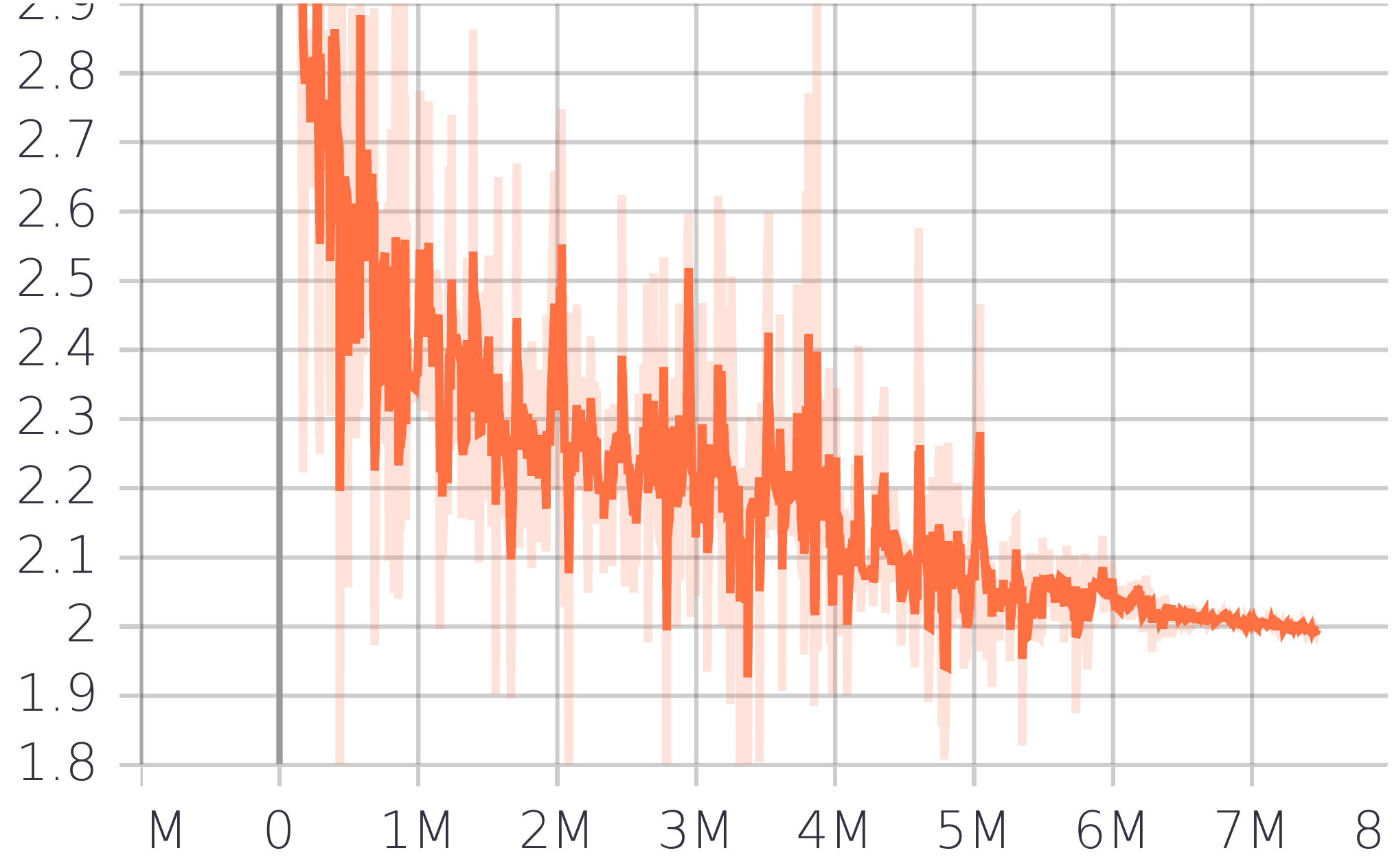}
         \caption{$\sigma^2$ - \emph{I-C} task}
         \label{fig:sigma_ic}
     \end{subfigure}
 \hfill
     \begin{subfigure}[b]{0.23  \textwidth}
         \centering
         \includegraphics[width=\textwidth]{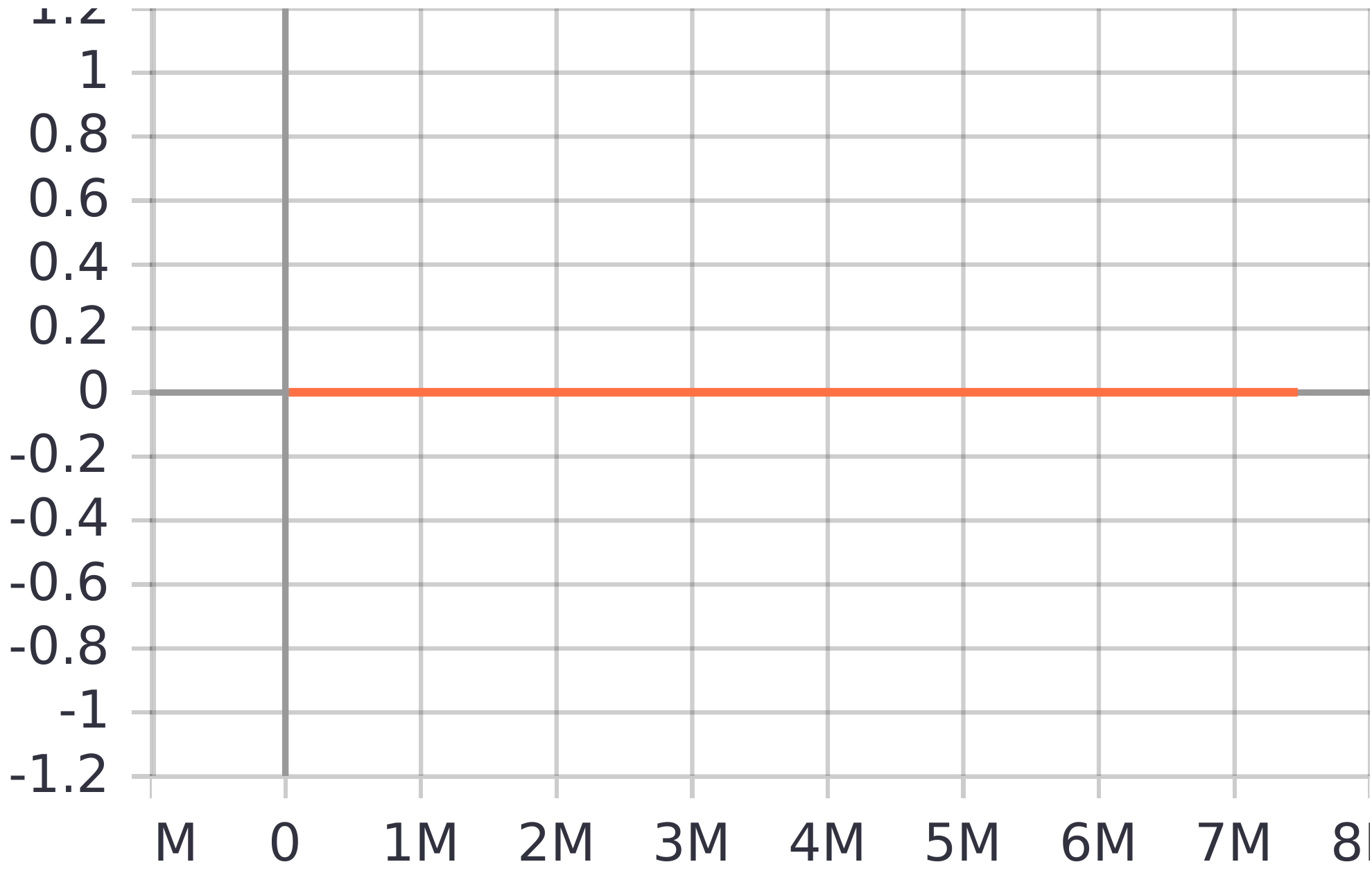}
         \caption{$\sigma^2$ - \emph{I} task}
         \label{fig:sigma_i}
     \end{subfigure}
 
  \begin{subfigure}[b]{0.23 \textwidth}
         \centering
         \includegraphics[width=\textwidth]{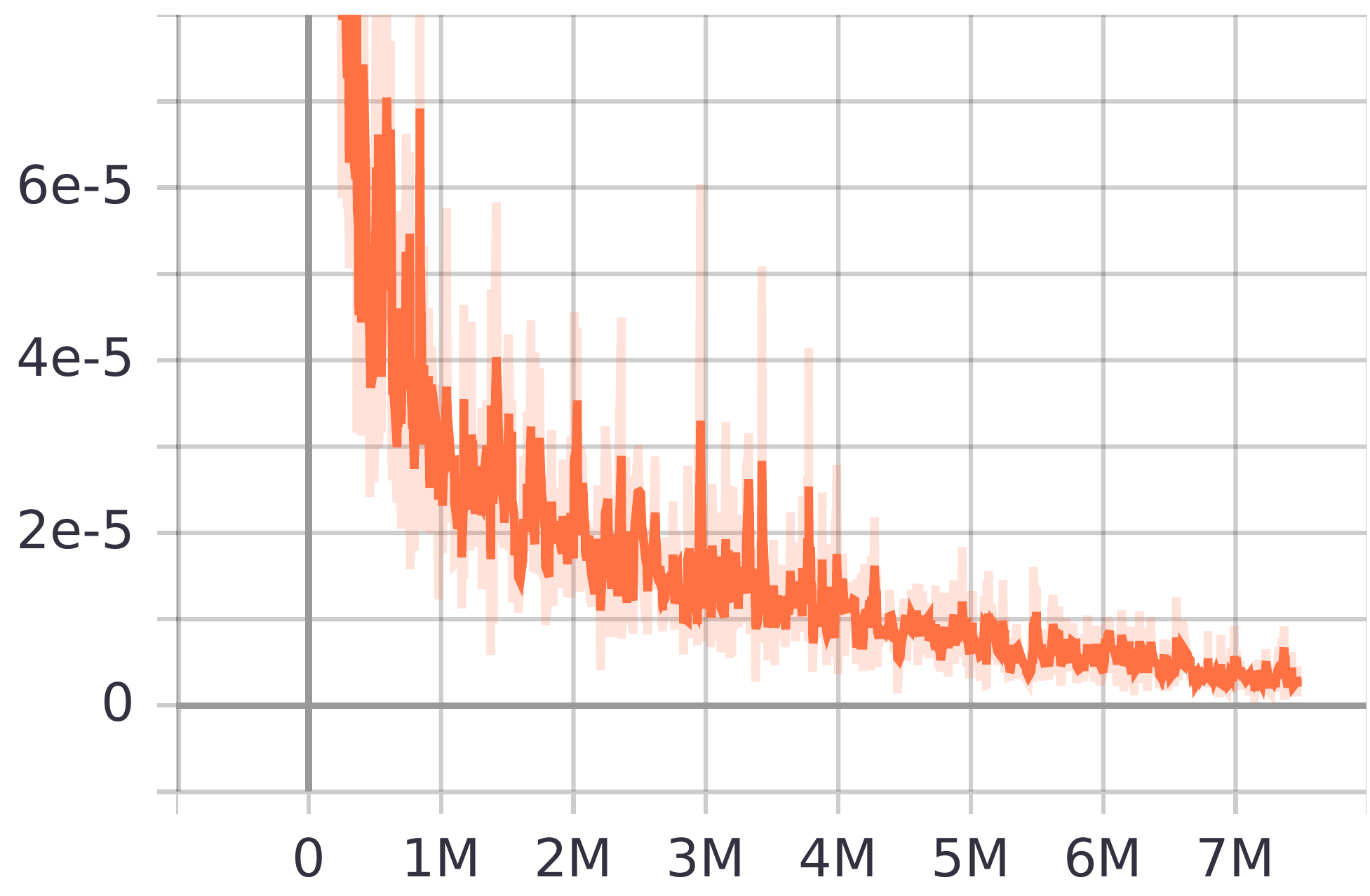}
         \caption{Loss - \emph{I-C} task}
         \label{fig:loss_ic}
     \end{subfigure}
  \hfill  
     \begin{subfigure}[b]{0.23 \textwidth}
         \centering
         \includegraphics[width=\textwidth]{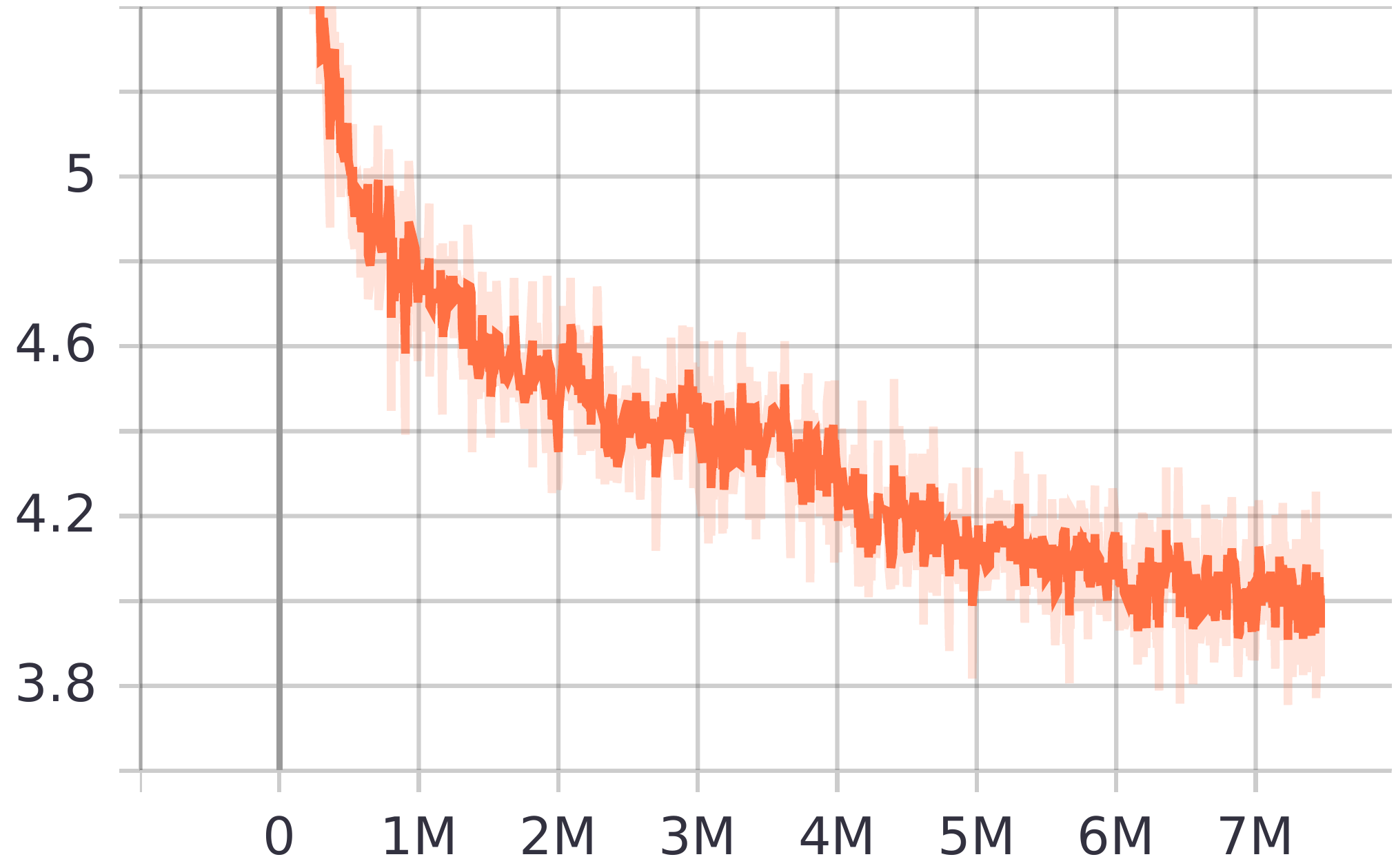} 
         \caption{Loss - \emph{I} task}
         \label{fig:loss_i}
     \end{subfigure}   
     
        \caption{$\sigma^2$ and losses of the \emph{instance - category (I-C)}) inter-view task and the \emph{instance (I)} intra-view task. Note that the horizontal axis is the number of training steps and the loss weight of each task is $1/ \sigma^2$.}
        \label{fig:four graphs}
\end{figure}

\begin{table}[t]
\caption{Performance (HitRate@50) of different loss weighting strategies.}
\label{exp:loss_weight}
\begin{tabular}{l|ccc}
\toprule
 &M2GRL$_{uniform}$ & M2GRL$_{static}$  & M2GRL$_{adapative}$ \\ \midrule
 Movielens& 42.24\%&42.50\%  & 42.62\% \\ \midrule
Taobao &33.08\% &33.11\%  & 33.24\% \\ \bottomrule
\end{tabular}
\end{table}

\subsection{Analysis of Loss Weighting}
To evaluate the influence of the adaptive loss weighting mechanism (denoted by \method{}$_{adaptation}$), we experiment with two variants of \method{}: \method{}$_{uniform}$ with uniform weights and \method{}$_{static}$ with manually assigned weights. We perform a grid search and find the optimal weights are approximately proportional to the distribution of the training data. Concretely, we set $0.05$ as the loss weight for the instance-view task, and $1$ for the other tasks. From Table~\ref{exp:loss_weight}, we can see that \method{}$_{adaptation}$ outperforms others on two datasets, showing the superiority of the adaptive loss weighting mechanism. Note that \method{}$_{static}$ performs better than \method{}$_{uniform}$, but finding proper task weights is too expensive. It's almost impossible to search task weights manually for a daily updated industrial algorithm due to computation cost concerns. 

Besides, in Figure~\ref{fig:four graphs} we display the uncertainty variable $\sigma^2$ (loss weight is $1/ \sigma^2$) and the losses of two tasks of \method{}$_{adaptation}$: the \emph{I-C} inter-view task and the \emph{I} intra-view task. Note that the loss weights are unnormalized and we set a threshold ($0.05$) to avoid weight explosion. The procedure is similar to gradient clipping: if updated weights are below the threshold, we set the weights to the threshold. 
First, we can see that in Figure~\ref{fig:sigma_ic} $\sigma^2$ decreases as the training step increases, and in Figure~\ref{fig:sigma_i} $\sigma^2$ is below the threshold so the loss weight is set to $1/0.05=20$. Hence, compared with the \emph{I} task, the loss weight of the \emph{I-C} task increases as training goes on. Second, in Figure~\ref{fig:loss_ic} and \ref{fig:loss_i}, the loss of each task decreases steadily as the training steps increase, but the loss of the \emph{I-C} task is much smaller and thus is prone to be interfered by the \emph{I} task. The adaptive weight mechanism helps to alleviate this issue and guarantee the convergence of each task.

\subsection{Visualization}
\begin{figure}[t]
     \centering
     \begin{subfigure}[b]{0.23 \textwidth}
         \centering
         \includegraphics[width=\textwidth]{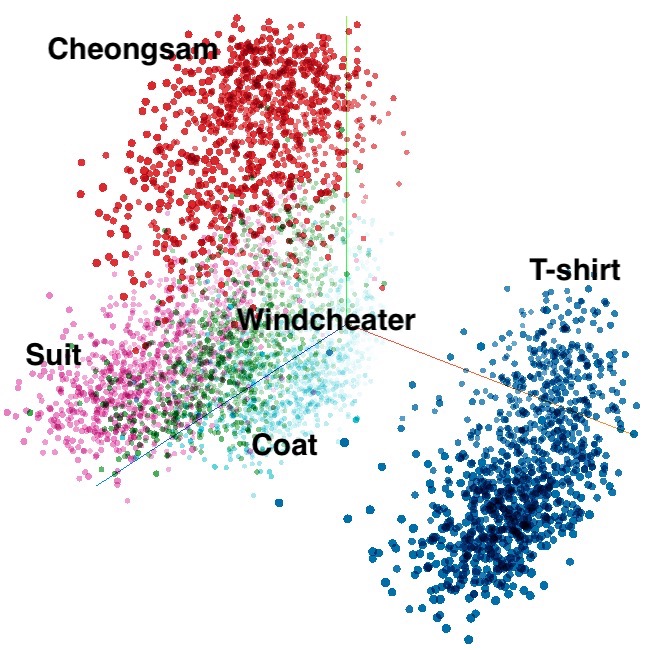}
         \caption{Instance-view embeddings.}
         \label{fig:case_study1}
     \end{subfigure}
 \hfill
     \begin{subfigure}[b]{0.23  \textwidth}
         \centering
         \includegraphics[width=\textwidth]{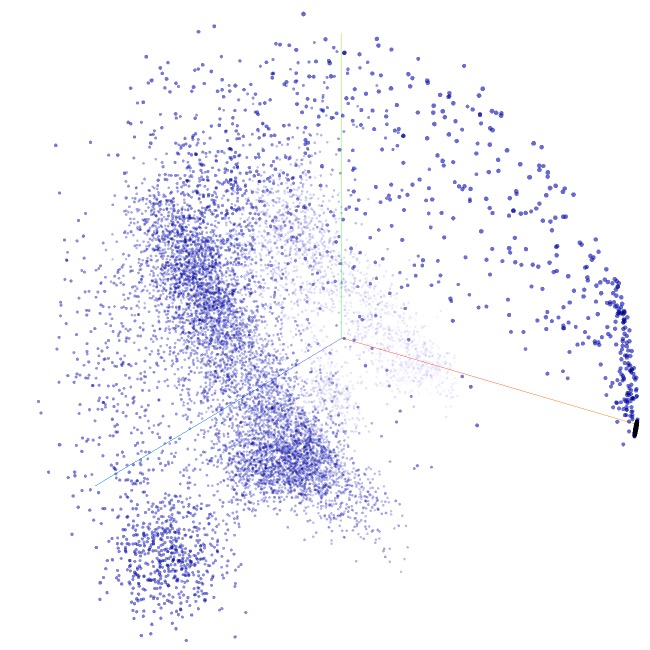}
         \caption{Category-view embeddings.}
        \label{fig:case_study2}
     \end{subfigure} 
     
        \caption{Visualization of the learned embeddings.}
        \label{fig:case_study}
\end{figure}

In this section, we present some visualization results of the learned representations of real-world cases in Taobao to further illustrate the effectiveness of \method{}. We visualize the instance embeddings of items via the Embedding Projector\footnote{https://projector.tensorflow.org/}, a visualization tool provided by Tensorflow. We apply principal component analysis (PCA) and draw the results in Figure \ref{fig:case_study}. In Figure \ref{fig:case_study1}, we choose five cloth-related categories (i.e., T-shirt, coat, suit, windcheater, and cheongsam), and display $1000$ randomly chosen items for each category. Here one color indicates one category of items. We can see that items of one category are clustered in the feature space and distances between different categories are different. There are two interesting observations: the T-shirt cluster is far away from other clusters, and the clusters of windcheaters and coats have many overlaps. The far distance can be explained that T-shirts have short sleeve and sometimes are treated as undergarments, while items of other four categories are outer garments. The reason for the overlaps is that windcheaters and coats are similar in function and windcheaters are annotated as coats in some shops in Taobao. Although the learned embeddings of these these cloth-related items should be similar because users may click them within a short time, \method{} can still well capture their category relations, which demonstrates the effectiveness of the learned embeddings of \method{} with multi-view data. In Figure \ref{fig:case_study2}, we also display the embeddings of $12,000$ categories in the Taobao dataset. We can find that theses categories are distributed unevenly in the embedding space. Some categories are distributed closely and tend to form clusters while some are scattered in the space, which further reveals \method{}'s ability to model category relations.

\section{Beyond recommendation accuracy}\label{sec:metric}
High recommendation accuracy is not a silver bullet to a successful recommender system; there are many other factors need to be considered to evaluate broader aspects of user satisfaction, such as diversity, utility, and user coverage. We argue that utilizing multiple representations is a promising direction for various recommendation tasks with different focus. Below we describe a use case of diversity recommendation in Taobao with multiple represenrations.

\begin{figure}[t]
\centering 
\includegraphics[width=0.4\textwidth]{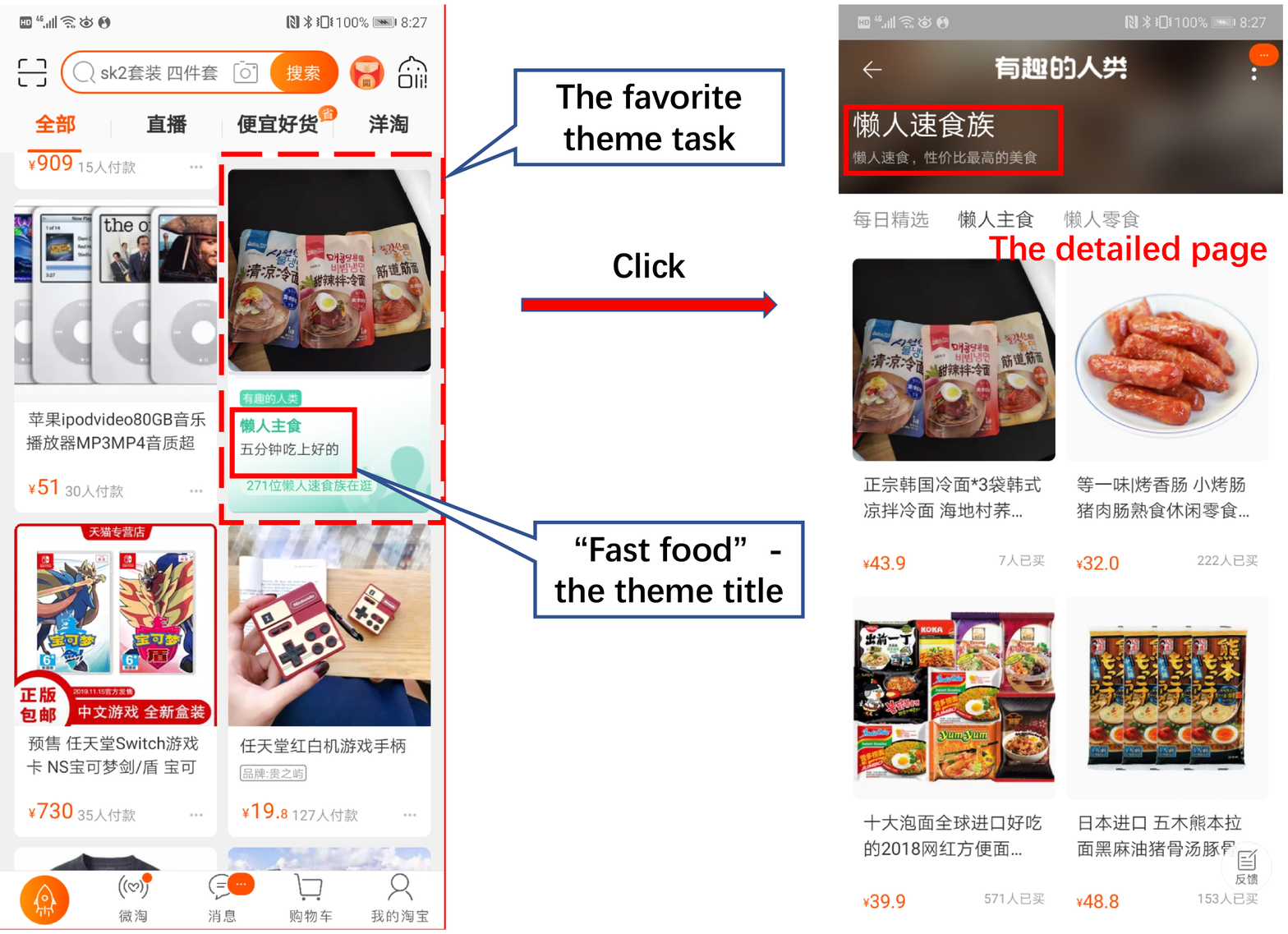}
\caption{An example of the ``favorite theme'' recommendation task.}
\label{fig:diversity_case}
\end{figure}

\subsection{Diversity Recommendation with Multiple Representations - A Downstream Use Case}\label{downstream_task}
In Taobao, recommendation diversity is an important metric for users' long-term experience. Customers will get bored if the system keeps recommending similar items of one category (e.g., skirts with different colors). Diversity recommendation in Taobao refers to recommending items whose categories did not appear in users' recent behavior logs (usually within two weeks). In traditional recommendation algorithms, items of the same category tend to have a high similarity score, which contradicts the goal of diversity recommendation. We argue that the multiple representations learned by \method{} can be utilized for diversity recommendation in a principled way. Heuristically, diversity recommendation is about finding two items that are relatively close in the instance-view embedding space but far away in the category-view embedding space. Thus, we propose a simple multi-view metric learning model. We use two representations, one is $e_i$ from the instance-view embedding space and the other is $e_{ic}=\sigma(W_{ic} \cdot e_{i})$ from the instance-category relational embedding space. For an item pair $(a,b)$, we use two metric matrices (i.e., $M_i$ and $M_{ic}$) to measure their distance:
\begin{equation}\label{eq:distance}
\begin{split} 
&d_i(a,b) = (e_{i}^{a} - e_{i}^{b})^T M_i (e_{i}^{a} - e_{i}^{b})\\
&d_{ic}(a,b) = (e_{ic}^{a} - e_{ic}^{b})^T M_{ic} (e_{ic}^{a} - e_{ic}^{b})\\
&d(a,b)  = d_i(a,b)  + d_{ic}(a,b)  \\
\end{split}
\end{equation}

Then, we use contrastive loss~\cite{hadsell2006dimensionality} to learn the metric matrices, which is
 \begin{equation}\label{eq:loss_metric}
L_{metric} = \frac{1}{2N} \sum_{n=1}^{N} yd^2+(1-y)\max(margin-d,0)^2,
\end{equation}
where $N$ is the number of training samples, $y$ is the label indicating whether the item pair is positive or negative, and $margin$ is a threshold to constrain the distance of negative samples. 

\begin{figure}[t]
\centering 
\includegraphics[height=4cm]{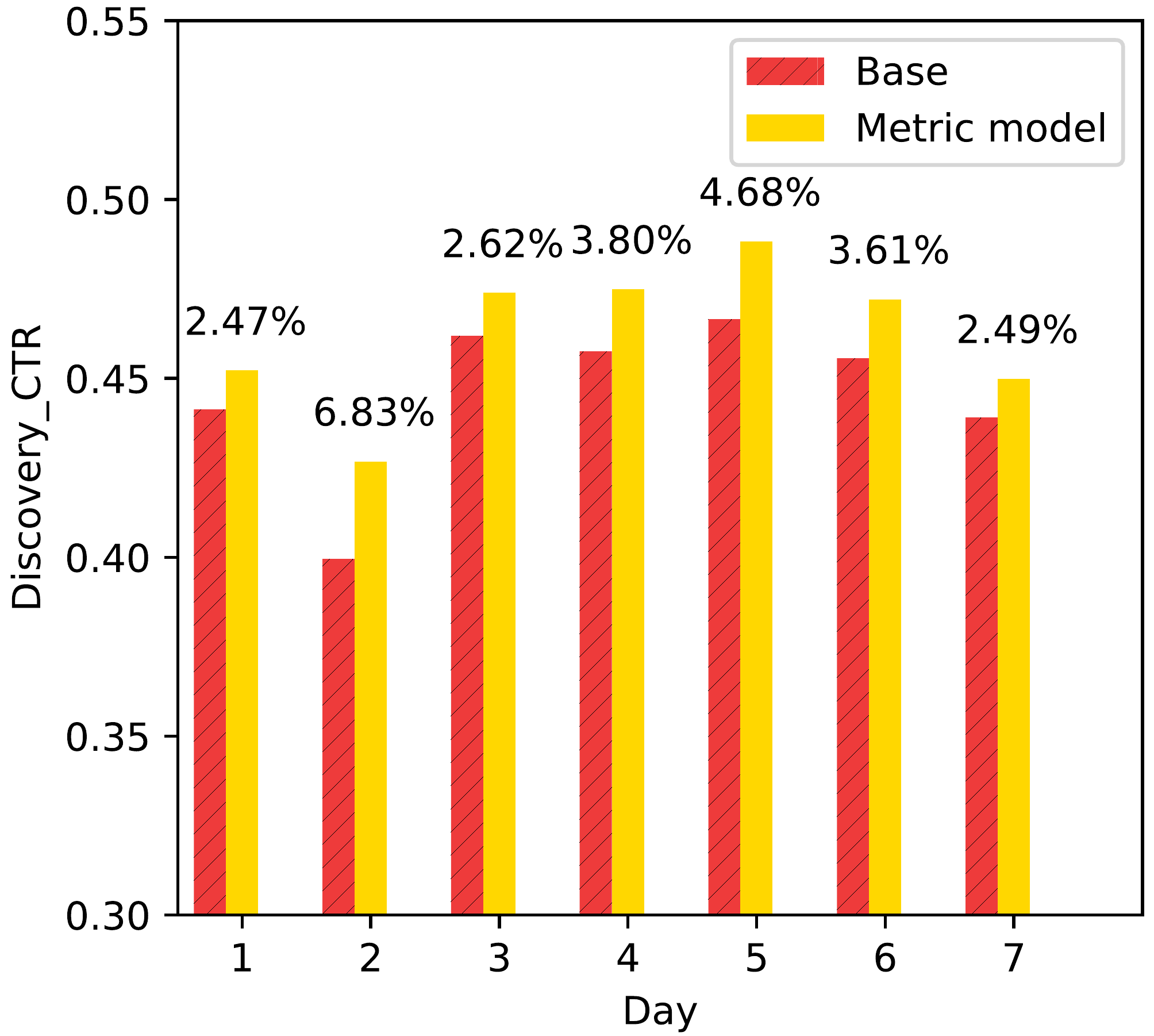}
\caption{The \emph{`Discovery CTR'} of the metric model and the base model. The numbers above histograms are improvements of the metric model against the base model.}
\label{fig:discovery_ctr}
\end{figure}

The above model is deployed at Taobao for the task of ``favorite theme''. As shown in Figure~\ref{fig:diversity_case}, the ``favorite theme'' task first shows a green card with an item and a theme (short text) to users. If users click that card, the screen will jump to the detailed recommendation page showing a subset of items within the clicked theme. The goal of the ``favorite theme'' task is to increase the diversity of recommendations and broaden the scope of users' knowledge of items, so the main challenge is to choose proper items with proper categories for the green cards. We used an empirical metric coined \emph{`discovery CTR'}, which is defined as $\frac{\#\text{clicks in new categories}}{\#\text{all clicks}}$, to evaluate diversity recommendation. The new categories denote categories that a user did not interact with in past $15$ days. We report the results of a seven-day online A/B test in Figure \ref{fig:discovery_ctr}. The base model directly uses representations of \method{} for candidate generation, and we can see that \method{}+metric model achieves an average $3.79\%$ relative improvement in \emph{`discovery CTR'}. The results demonstrate the usefulness of the learned multiple representations in the ``favorite theme'' task. \method{}+metric model is similar in some sense to the pretrain+finetune mode. Since there are various recommendation tasks with different goals, we argue that this methodology is promising for industrial recommender systems and the multiple representations generated by \method{} can provide more aspects of information to downstream tasks. 

\section{Conclusions}
We propose \method{}, a multi-task multi-view graph representation learning framework for web-scale recommender systems. \method{} supports unlimited number of views of data and can be distributedly deployed to handle billion-scale data samples. We also design an adaptive weight tuning mechanism to make \method{} more robust and scalable. We deployed \method{} at Taobao, and the offline experiments and online A/B test both shown the superiority of \method{} over other competitive methods. Besides, we explore a multi-view metric model for diversity recommendation with two representations learned by \method{}, which shows promising results. As there are many recommendation tasks with different focus in industry, e.g., tag recommendation and recommendation with explanations, we argue that the multiple representations generated by \method{} can be further utilized to tackle these tasks in the future.   
\section{Acknowledgments}  
X. Wu was supported by the grant of P0001175 funded by PolyU.
\bibliographystyle{ACM-Reference-Format}
\bibliography{multi-view19-1}


\appendix
\section{Appendix}
 
 \balance
\subsection{Data preprocessing}
The data preprocessing of the Taobao dataset is described in Sec 3.1 and 5.1. Here we introudce the data preprocessing of the Movielens dataset. For Movielens, we treat tag of movie as movie's category information and construct single-view graphs similarly. Note that Movielens does not have shop information, so we only use two graphs in \method{} for Movielens. The main preprocessing phases are as follows:
\begin{itemize}
    \item \emph{Data Clean}. In Movielens, user rating ranges from $1$ to $5$. It is commonly assumed that ratings below 3 indicate that users dislike the corresponding films. So we delete the ratings that are below $3$. As user only have to rate one film once, we do not have duplicate ratings. Note that this operation is also applied in graph construction. 
    \item \emph{Session Split}. We use timestamps to split user behaviors into sessions. We noticed that users' sequential behaviors are sparser in Movielens; the time lag between two consecutive rated films may be longer than a few years. To set a proper length of data sessions for \method{}, we split a session into two sessions if there is a one-year idle period. Moreover, if a session is too long (larger than $50$), we split the session into two sessions.
\end{itemize}
After preprocessing, we construct the item graph and category graph. For the item graph, we assume that two items are connected if they occur consecutively in one user's behavior history.

\subsection{Running Environment}
 In offline experiments, all the models are tested on MaxCompute (https://cn.aliyun.com/product/odps), which is a data processing and computing platform in Alibaba. To be fair, all the models share the same training and testing datasets. For Taobao, we use 10 parameter severs (PSs) and 50 GPU workers. For Movielens, we use 1 PS and 1 GPU worker. Computing resources: 1) CPU: each PS has $20$ cores and $30GB$ memory, and each worker has $8$ cores and $10GB$ memory. 2) GPU: Tesla P100-PCIE-16GB.


\subsection{Algorithm settings and parameters}
In this section, we describe some algorithm settings and parameters related to the experiments. EGES$_{asy}$, DeepWalk, Node2Vec, Line, and Graphsage are tested on the Aligraph, a graph representation learning paltform in Alibaba.
We implement distributed versions of GRU4Rec and YouTube-DNN to support the big data. 
We mainly use the original parameters for the baselines. Below we list some important parameters of each algorithm. If not specifically noted, the batch size is set to $256$.
\begin{itemize}
\item \method{}: we set batch size to $2048$, epoch num to $10$, and the number of negative samples to $10$. Adam optimizer with learning rate 0.01 is used to update parameters, and gradient clipping is adopted by scaling gradients when the norm exceeds a threshold of 1.
\item EGES$_{asy}$:  The settings of EGES$_{asy}$ are the same as those of \method{}.

\item DeepWalk: We set vector dimen to $64$, negtive sampling num to $7$, batch size to $512$, epoch num to $100$, and random walk length to $5$. We have tuned the walk length to $18$ and set window size to $9$ in order to match EGES$_{asy}$ and \method{}, but the performance is worse than the current setting.
\item Node2Vec: The settings of Node2Vec are the same as those of DeepWalk except that $q=0.5$, $p=1$.

\item LINE: We choose to preserve the second-order proximity in the graph. The number of negative samples is $10$, and the number of vector dimension is set to $64$.

\item GraphSage: We train GraphSage in an unsupervised manner and set batch size to $256$ and dropout to $0$. We choose a two-layer structure, and dimensions for the first layer and the second layer are $20$ and $10$ respectively. Due to engineering concerns, we did not use the attribute information of items.

\item GRU4rec: We set sequence length to $30$, epoch num to $10$, GRU layer number to $1$, bucket size to $10$, and the number of samples to $1000$.

\item Youtube-DNN: We set sequence length to $30$, number of units to $64$, embedding size to $64$, and the number of samples to $1000$. In our experiments, vectors of items and users are concatenated and fed to a multi-layer perceptron network.

\end{itemize}
 
 Two used metrics are described here. 
 HitRate@K represents the proportion of positive test cases $n_{hit}$ which have relevant (correctly recommended) items among top K of the ranking list to all test cases $N$, defined as
    \begin{displaymath}
    \mbox{HitRate@K} = \frac{n_{hit}}{N}.
    \end{displaymath}
    
    F1@K is calculated by
 \begin{displaymath}
 \frac{2 \times Precision@K \times Recall@K}{ Precision@K + Recall@K }.
   \end{displaymath}  

\subsection{Code Release}
A demo code of M2GRL is released to \href{https://github.com/99731/M2GRL}{\color{blue}{https://github.com/99731/M2GRL}}.

\end{document}